\documentclass[12pt]{article}

\usepackage{graphicx}

\begin{document}

\title{Ferroelectricity of Li-doped silver niobate (Ag,Li)NbO$_3$}

\author{Desheng Fu$^{1}$, Makoto Endo$^{2}$, Hiroki Taniguchi$^{2}$,\\
Tomoyasu Taniyama$^{2}$,Mitsuru Itoh$^{2}$, and Shin-ya
Koshihara$^{3,4}$\\$^1$Division of Global Research Leaders, Shizuoka
University,\\ Johoku 3-5-1, Naka-ku, Hamamatsu 432-8561,
 Japan \\
 $^2$Materials and Structures Laboratory,
Tokyo Institute of Technology,\\ 4259 Nagatsuta, Yokohama 226-8503,
Japan\\
$^3$Department of Materials Science, Tokyo Institute of Technology,
\\Meguro-ku, Tokyo
152-8551, Japan\\
$^4$CREST \& ERATO, Japan Science and Technology Agency (JST),\\ 3-5
Sanbanchou, Chiyoda-ku, Tokyo 102-0075,
Japan\\E-mail:ddsfu@ipc.shizuoka.ac.jp}


\maketitle

\begin{abstract}
Phase evolution in (Ag$_{1-x}$Li$_x$)NbO$_3$ (ALN) solid solution
was investigated by X-ray diffraction technique, dielectric and
polarization measurements. It is shown that small substitution of Ag
with Li gives rise to an  orthorhombic-rhombohedral structural
transformation in ABO$_3$-perovskite silver niobate at room
temperature. Structural refinements indicate that both  A- and
B-site displacements contribute to the spontaneous polarization of
the ferroelectric  phase with  symmetry  $R3c$. Increasing
Li-concentration enhances the ferroelectric rhombohedral distortion,
resulting in the increase of the para-ferroelectric phase transition
temperature and the polarization of the solid solutions.
\end{abstract}


\section{Introduction}
  Ferroelectric materials, which have high dielectric constant and
large spontaneous polarization, offer various applications ranging
from electronic devices, telecommunication, medical imaging and
ultrasonic devices, to ferroelectric memories
\cite{Lines,Jaffe,Scott}. The prototype ferroelectrics include
BaTiO$_3$ and PbTiO$_3$: the former is generally used for the
fabrication of the capacitive devices while the latter is a base
material for the piezoelectric devices. For BaTiO$_3$, the problems
that  hamper its technological applications are the discontinuity of
physical properties around room temperature due to a structural
phase transition  and the relatively-low ferroelectric phase
transition $T_{\rm c}^{\rm FE}$ ($\sim 400 K$) \cite{fu0}. On the
other hand, PbTiO$_3$-based materials show high $T_{\rm c}^{\rm FE}$
and high dielectric/piezoelectric responses, but contain hazardous
lead, which raises a serious environmental concern\cite{Saito}.
Currently, increasing effort is being made to develop lead-free
ferroelectric materials with high $T_{\rm c}^{\rm FE}$ and high
performances of physical properties
\cite{Saito,fu1,fu2,fu3,Bilc,Grinberg}.

  Ferroelectricity in the  ABO$_3$-perovskites is basically driven by the cooperative displacement of
B-site atoms filling the octahedral site. Modern investigations
further reveal that the strong hybridization between metal atom and
oxygen is essential for the origin of ferroelectricity in these
perovskite oxides \cite{Cohen}. In contrast to BaTiO$_3$ in which
polarization  is originated from the B-site atomic  displacement,
the lead-based materials generally show large A-site atomic
displacement in the dodecahedral cave in addition to the B-site
atomic  displacements\cite{Egami}. It is considered that the strong
covalency of Pb-O  effectively softens the short-range repulsive
interaction between atoms and favors Pb to move into an
off-centering position in the perovskite structure of the lead
compounds.\cite{Cohen,Kuroiwa} It is generally accepted that such
hybridization characteristic is crucial for the high piezoelectric
performance of the lead-based materials. Apparently, it also
outlines a basic direction toward searching  novel element to
replace the hazardous Pb in the perovskite oxides.  Ag seems to be a
promising candidate because it has been shown theoretically that Ag
has a strong hybridization with oxygen in the perovskite compounds
\cite{Grinberg,Grinberg2,Kato}. From the recent structural data
\cite{Sciau}, we can evaluate the bond lengths of Ag-O in AgNbO$_3$,
and find that  the shortest  bond length of Ag-O is indeed
significantly less than the sum of silver and oxygen ionic radii,
supporting the theoretical prediction of the strong covalent bonding
in silver perovskites.  It is naturally expected that A-site driven
polarization may be realized in the silver perovskites. Actually,
experimental investigations reveal that a ferroelectric state with
exceptionally high polarization (52 $\mu$C/cm$^2$ for polycrystal
sample) can be induced  in  AgNbO$_3$ through the application of a
high electric field.\cite{fu4} Thus both theoretical and
experimental findings  suggest that Ag posses the essential features
of Pb to cause a strong ferroelectric distortion in the niobate
perovskite, and good ferroelectrics  are highly expected in the
silver perovskites.

In addition to the the application of a high electric field,
preliminary investigations indicate that small substitution of Ag
with K or Li can change the structure of AgNbO$_3$, in which oxygen
octahedral tilting due to the low tolerance
factor\cite{Sciau,Francobe} occurs and  hampers the ferroelectric
distortion, into a ferroelectric structure with large polarization
and piezoelectric response\cite{fu5,fu6}. Currently,  there is
rather large disagreement in literature about the crystal structure
of AgNbO$_3$.\cite{Levin,Pawelczyk,Petzelt} Also, the reports on the
structure of its solid solutions are very limited and sometime
confused. \cite{Nalbandyan,Sakabe,Wada}. The origin of its
ferroelectricity is still lack of understanding. In this study, we
carried out a systematical investigation on the effects of Li
substitution on the ferroelectricity of AgNbO$_3$. We show that the
evolution of ferroelectricity is closely related to the rhombohedral
distortion caused by the Li-substitution. The present investigations
may improve our understanding of the ferroelectricity in the
silver-based perovskites.

\section{Experimental}
    Ag$_{1-x}$Li$_x$NbO$_3$ (ALN) polycrystals were prepared by a solid
state reaction approach. Appropriate amounts with stoichiometric
ratio of  Ag$_2$O, Nb$_2$O$_5$, and Li$_2$CO$_3$ were mixed
homogeneously with ethanol and calcined at 1253 K for 6 hours in
O$_2$ atmosphere, followed by removing the powder out of the furnace
to allow a rapid cooling for preventing the phase separation. The
calcined powder was milled again and pressed to form pellets that
were sintered at 1323 K for 6 hours in O$_2$ atmosphere, followed by
a rapid cooling. Single crystal samples were obtained from the
calcined powder by a melt growth process described in a previous
report\cite{fu5}.

X-ray diffraction technique was used to determine the structure and
lattice parameters of ALN solid solutions. For electrical
measurements, the sintered pellets were polished and coated with the
Au electrodes. Dielectric measurements were performed in a
temperature range of 300 K $\sim$ 850 K  by using a Hewlett-Packard
Precision LCR meter (HP4284A) at an ac level of 1V/mm. Dielectric
($D-E$) loops were measured at room temperature with a ferroelectric
measurement system of aixACCT TF Analyzer 2000 equipmented with a
high voltage source of 10 kV. To prevent air breakdown at the high
field, samples were immersed in silicon oil during the measurements.

Inductively coupled plasma (ICP) spectrometry (HORIBA-JOBIN-YVON
ULTIMA2) was used to determine the chemical composition  of samples
with an error less than 0.2 mol\%. To prepare the solution for ICP
measurements, 2 mg of the same sample used for electrical
measurements was mixed with 10 ml sulfuric acid and 2g ammonium
sulfate, and dissolved at 503 K for 20 minutes with autoclave. ICP
analysis indicates that the deviation of Li concentration from the
starting material for ceramics samples is less than 1\%, while
relatively-large deviation is observed for the single crystals.

\section{Results and Discussions}

\subsection{X-ray diffraction patterns}
\begin{figure}[H]
\includegraphics[height=14cm]{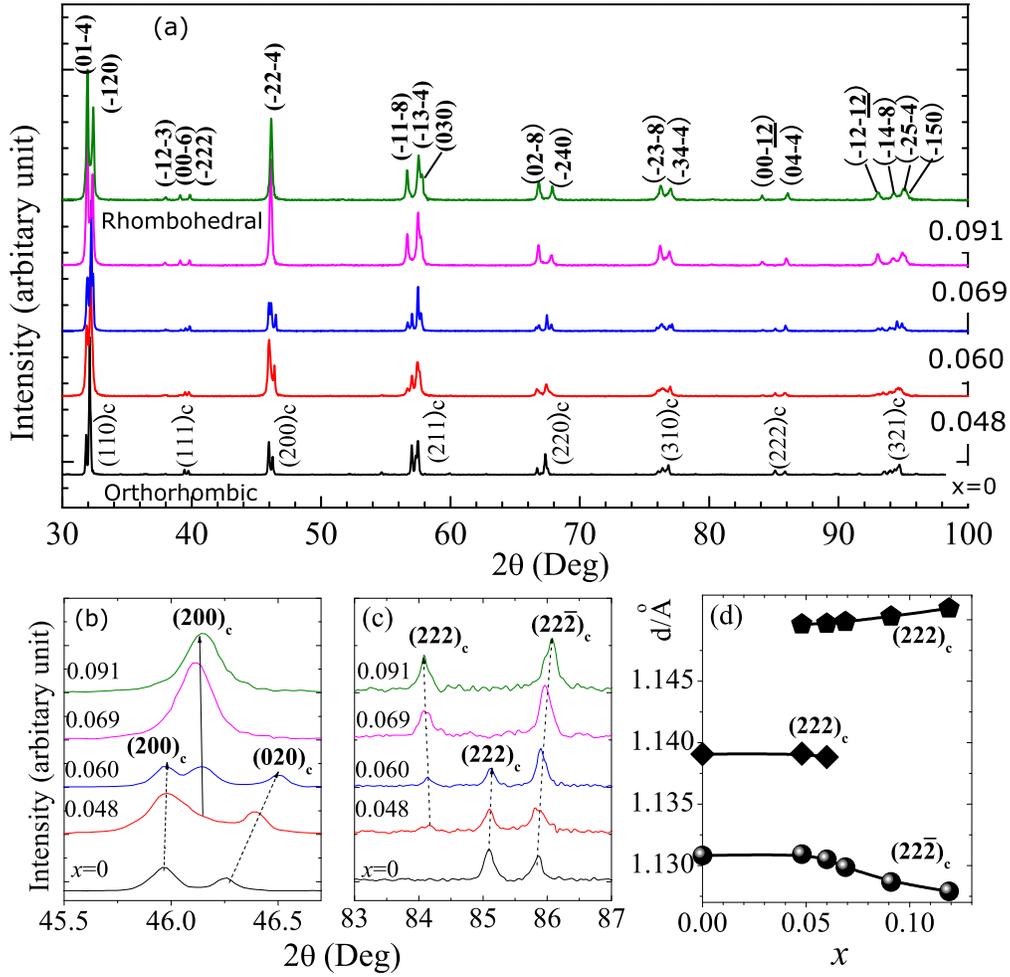}
\caption{\label{fig1}  (a) Powder X-ray diffraction patterns of
Ag$_{1-x}$Li$_x$NbO$_3$ solid solutions. The major diffraction peaks
are also indexed. For simplification, the plane indices of
pseudocubic structure (indicated by a subscript of {\it c}) are
given for the mother material AgNbO$_3$. (b)  and (c) show the
enlarged views of the diffraction peaks of (200)$_c$ and (222)$_c$,
respectively. Structure change can be clearly seen from the
splittings of these diffractions. (d) $d$-spacings vs concentration
for the (222)$_c$ planes. }
\end{figure}

  Figure \ref{fig1} shows the  X-ray diffraction patterns observed
at room temperature for several compositions of ALN powders. It is
immediately clear that there is a structural transformation as
incorporating Li into Ag site. The phase boundary is found to locate
at composition of $x_{\rm c}\approx0.05-0.06$. For $x<x_{\rm c}$,
the solid solution posses the orthorhombic structure of pure
AgNbO$_3$ \cite{Francobe,Nalbandyan}. When $x$ is larger than
$x_{\rm c}$, however, the solid solution transforms into a
rhombohedral structure\cite{Nalbandyan}. Such rhombohedral symmetry
can be unambiguous seen from the splittings of  200, 220, 222
pseudocubic reflections and supported from  structural refinements
described in the next section. This rhombohedral phase remains
unchanged for composition $x<\sim 0.12$. However, further increase
in Li-concentration will lead to the occurrence of second phase of
LiNbO$_3$ hexagonal pseudo-ilmenite-type.

\subsection{Structural refinements of the rhombohedral phase}
  The detailed structure of pure AgNbO$_3$ has been investigated by various techniques including x-ray
and neutron diffractions\cite{Sciau,Levin}. These structural
analyses predict centrosymmetric $Pbcm$ space group for the room
temperature phase. It should be pointed out that such
centrosymmetric space group is inconsistent with the experimental
results from dielectric, polarization and piezoelectric
measurements\cite{fu1,Kania1}, which indicate the existence of weak
ferroelectricity in AgNbO$_3$  at room temperature. Such discrepancy
is very likely due to the extremely-small atom displacement in the
structure, resulting in the difficulty of its detection by X-ray or
neutron diffractions. Actually, the reported X-ray or neutron
diffraction analyses could not distinguish between centrosymmetric
$Pbcm$ and noncentrosymmetric $Pbc2_1$ space groups\cite{Sciau}. The
exact determination of the space group of AgNbO$_3$ room temperature
phase goes beyond the scope of present investigation, and remains to
be addressed in future work.

\begin{figure}[H]
\includegraphics[height=12cm]{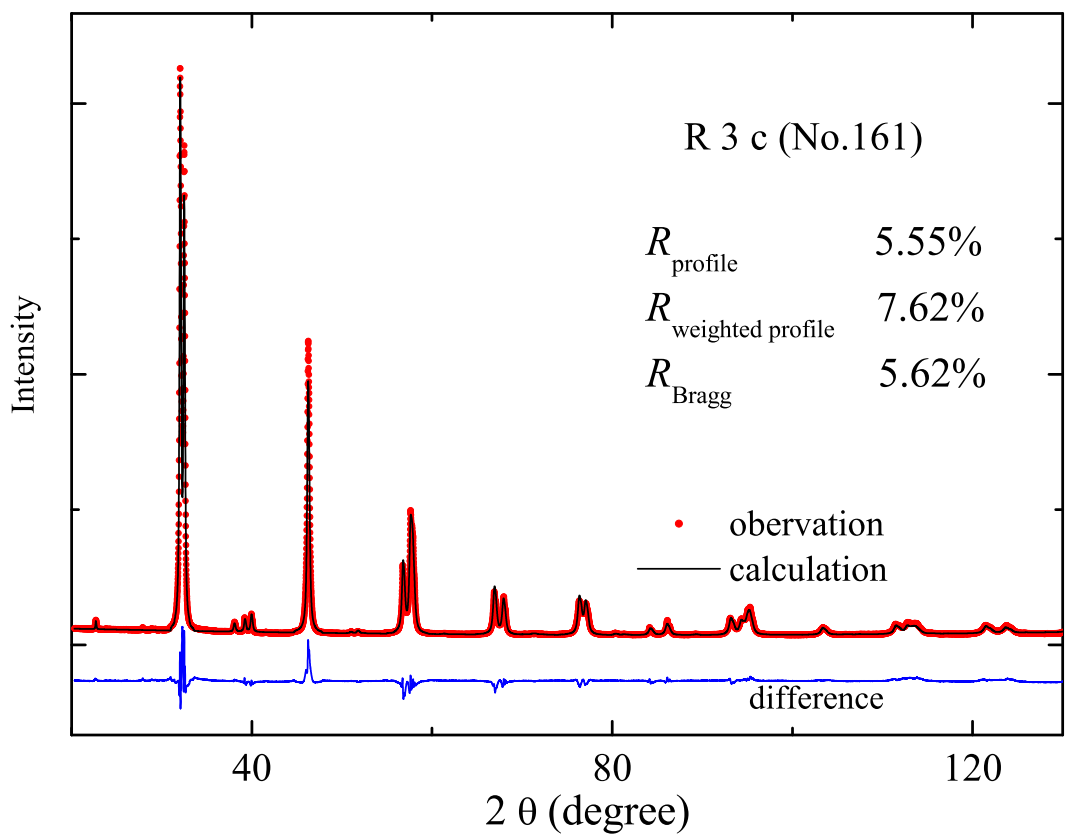}
\caption{\label{fig2} Observation, calculation and difference
profiles for Ag$_{1-x}$Li$_x$NbO$_3$ ($x=0.1$). Structural
refinements at room temperature were  performed with space group of
$R3c$(No.161). $R$-factors  were given in the figure.}
\end{figure}

On the other hand,  there is still a lack of knowledge of the
detailed structure of the ALN rhombohedral phase. In order to  gain
an insight into the ferroelectricity of this material, we then
carried out a preliminary analysis on the structure of the ALN
rhombohedral phase. To achieve such a purpose, we collected powder
X-ray diffraction data in the 2$\theta$ range of
$20^\circ-130^\circ$ at a step width of $0.01^\circ$ by using Bruker
AXS D8 ADVANCE powder diffractometer (X-ray source is the radiation
of Cu $K\alpha$). Measurements were performed at room temperature
with a powder sample of $x=0.10$. An intensity of about $6*10^5$
counts was measured for the strongest diffraction peak around
$2\theta\sim32.2^\circ$. The structural refinements were then
performed using the Rietveld refinement program X'Pert Plus. A
pseudo-Voigt profile function was used and the ratio of $K\alpha2$
and $K\alpha1$ was assumed as  0.5 during the refinements. We
obtained  refinements with reasonably-good $R$-factors for the space
group $R3c$. Figure \ref{fig2} shows the results of the refinements.
The fit between the observed and calculated profiles is quite
satisfactory. Table 1 lists the structural parameters for this phase
model. From this structure, we can see that this rhombohedral $R3c$
phase involves polar displacements of the Ag/Li, Nb, and O atoms
along the pseudocubic [111] direction. This is also supported by the
spontaneous polarization measurements\cite{fu5}, which show that the
polarization along [111] direction is larger than that along [001]
direction. This structural model supports our inference that
incorporation of Li into the Ag site favors the rhombohedral
ferroelectric distortion over the oxygen octahedral tilting in the
perovskite structure.

\subsection{Dielectric properties}
\begin{figure}
\includegraphics[height=16cm]{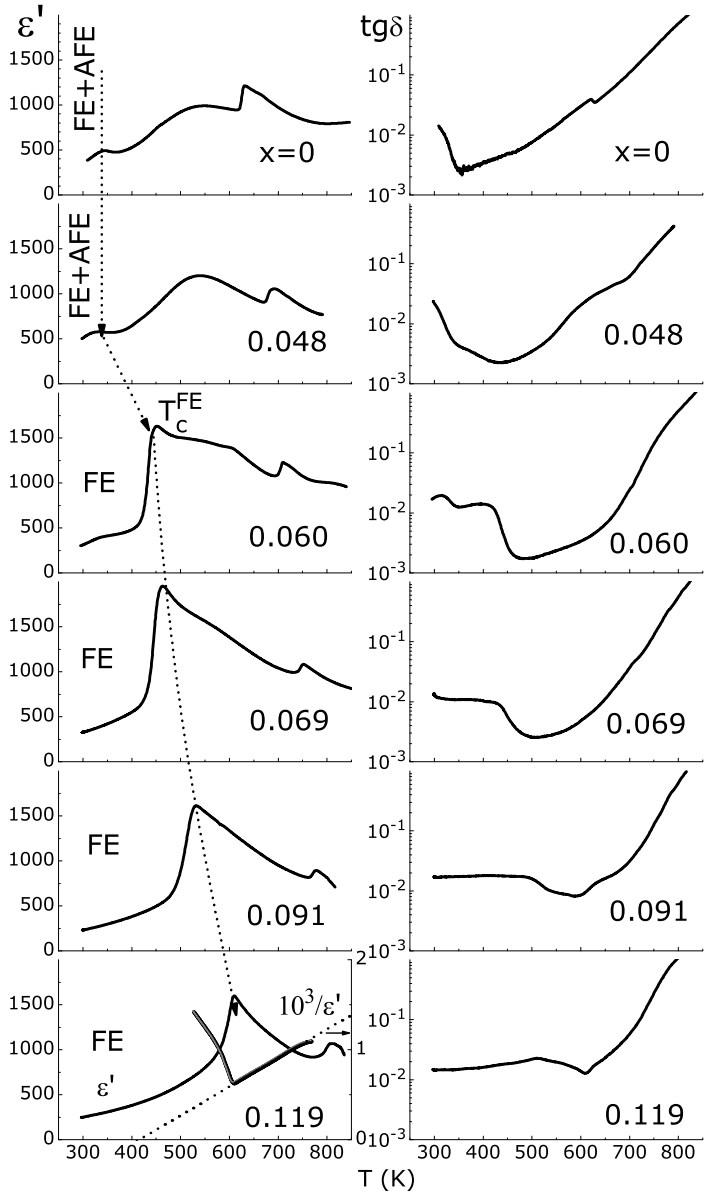}
\caption{\label{fig3}  Temperature dependence of dielectric constant
and loss. Data for $x=0$ are from Ref.\cite{fu4}. Reverse dielectric
constant is also given for the composition with $x=0.119$.}
\end{figure}

The phase transformation observed by X-ray diffraction technique is
also confirmed in the dielectric measurements(Fig. \ref{fig3}). For
$x<x_{\rm c}$, ALN exhibits the dielectric  behaviors similar to
that of pure AgNbO$_3$\cite{fu4}, indicating the same phase
transition sequences between them. Similar to AgNbO$_3$,  there are
four detectable anomalies of dielectric constant ($\varepsilon'$) in
the ALN solid solutions upon heating. For AgNbO$_3$, the cusp at
$\sim340$ K is interpreted to the disappearance of weak
ferroelectricity \cite{Kania1}. Following the cusp at $\sim340$ K,
there is a  broad peak at $\sim$540 K. It is suggested to be related
to the dynamics of antiparallel Nb displacements \cite{Kania2},
however, its origin remained to be clarified. A sharp jump at 624 K
can be well explained by an antiferroelectric phase transition
\cite{Sciau,Francobe,Pawelczyk}. A turning point following the
antiferroelectric phase transition can be seen at 668 K and is
basically caused by oxygen octahedral tilting transition
\cite{Sciau}. Generally, the variation in dielectric response is
small at transition point associating with the oxygen octahedral
tilting transition.

 In sharp contrast to composition with $x<x_{\rm c}$ , ALN solid solution  with $x>x_{\rm c}$
shows different temperature behaviors of the dielectric constant. A
Curie-Weiss-type ferroelectric phase transition occurs at $T=T_{\rm
c}^{\rm FE}$, and $T_{\rm c}^{\rm FE}$ increases with the increase
of the Li-concentration. Curie-Weiss constant is evaluated to be
$C=5.4\cdot10^{5}$ K for $x=0.119$. There is another dielectric
anomaly after the ferroelectric phase transition. Polarization
measurements indicate that this is a nonpolar phase. X-ray
diffraction measurements by Sakabe et al.\cite{Sakabe} suggest that
it is belong to  monoclinic phase. Apparently, further works are
necessary to understand structural change  occurring in the high
temperature range.

\subsection{Spontaneous polarization}
\begin{figure}
\includegraphics[width=12cm]{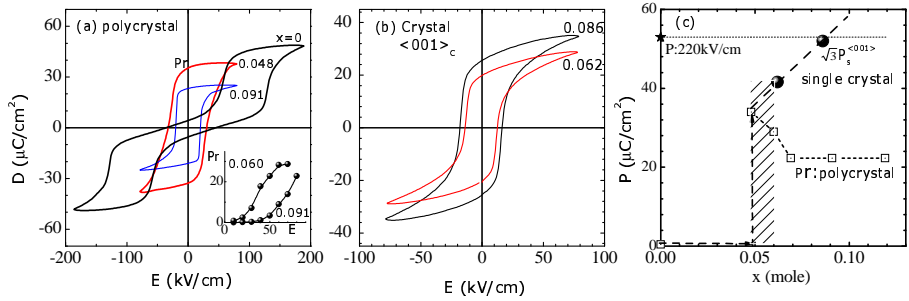}
\caption{\label{fig4}  $D-E$ hysteresis loops for (a) ceramics and
(b) $\langle001\rangle_{\rm c}$-oriented crystals. Inset shows the
electric field dependence of $P_{\rm r}$ for ceramics with $x=0.060$
and 0.091. (c) Composition dependence of remanent polarization
(square) measured at 80kV/cm for ceramics, and spontaneous
polarization (black circle) of single crystals; Shaded region
indicates the phase boundary between orthorhombic and rhombohedral
structures; Star ($\star$) and dash line indicate the polarization
value obtained at 220 kV/cm for AgNbO$_3$ ceramics
samples\cite{fu4}.}

\end{figure}

   In order to observe the evolution of the ferroelectricity in
 the ALN solid solutions, we also carried out measurements on $D-E$
 loops for the ceramics samples and two single crystals with $(001)_c$
 crystal face at room temperature. The results are summarized in
 Fig.\ref{fig4}. For pure AgNbO$_3$, only slim loop can be observed
 when electric field is lower than 110 kv/cm, but double hysteresis
 loops with extremely-large polarization can be clearly seen at higher field\cite{fu4}. Around the phase boundary,
 the double hysteresis loop changes into a normal ferroelectric loop, clearly indicating that the incorporation of Li into the Ag-site effectively unlocks the strong local polarization in
AgNbO$_3$. The remanent polarization $P_{\rm r}$ measured at  80
kV/cm for ALN ceramics are given in Fig. \ref{fig4}(c). Again, we
can see a sharp jump of  $P_{\rm r}$ at phase boundary, indicating
that a ferroelectric state with large polarization is established in
the rhombohedral phase.  All ceramics samples show $P_{\rm r}$ value
comparable to that of BaTiO$_3$ single crystal (26 $\mu$C/cm$^2$)
\cite{Shiozaki}. Moreover, the polarization in ALN solid solution is
very stable after switching. This is evident from the ideal square
shape of the loop. Large $P_{\rm r}$ value
 and ideal bistable polarization state of ALN ceramics may be
interesting  for the  non-volatile ferroelectric memory applications
because memory devices constructed generally from PZT or
Bi-layer-structure ferroelectrics suffer the poor retention of
stored information or degradation of performance due to the
polarization destabilization or rather small remanent polarization
\cite{Araujo}.

Although we observed a slight decrease in $P_{\rm r}$ with Li
concentration ($x>x_{\rm c}$) in ceramics samples, it can  be
reasonably accounted for in terms of  an insufficient poling field.
As shown in inset of Fig. \ref{fig4}(a), a saturation value of
$P_{\rm r}$ is not available for the  sample of $x=0.091$ at the
present applied field. Obviously, higher electric field are
necessary to obtain a saturation state.

In order to examine the effect of Li-substitution on spontaneous
polarization, single crystals with natural facet of pseudocubic
(001)$_{\rm c}$ were then used to obtain the $D-E$ loops. Fig.
\ref{fig4}(b) shows that  Li-substitution enhances the spontaneous
polarization of ALN with the rhombohedral structure. Because the
polar axis is along $\langle111\rangle_{\rm c}$ in the rhombohedral
phase as described above and shown in Ref.\cite{fu5}, the
spontaneous polarization is then estimated from the relationship
$P_{\rm s}=\sqrt{3}P_{\rm s}^{<001>}$, and is found to increase with
increasing Li-concentration as shown in Fig. \ref{fig4}(c).

\section{Discussion}

\subsection{Relationship between ferroelectricity and rhombohedral distortion}

\begin{figure}
\includegraphics[height=10cm]{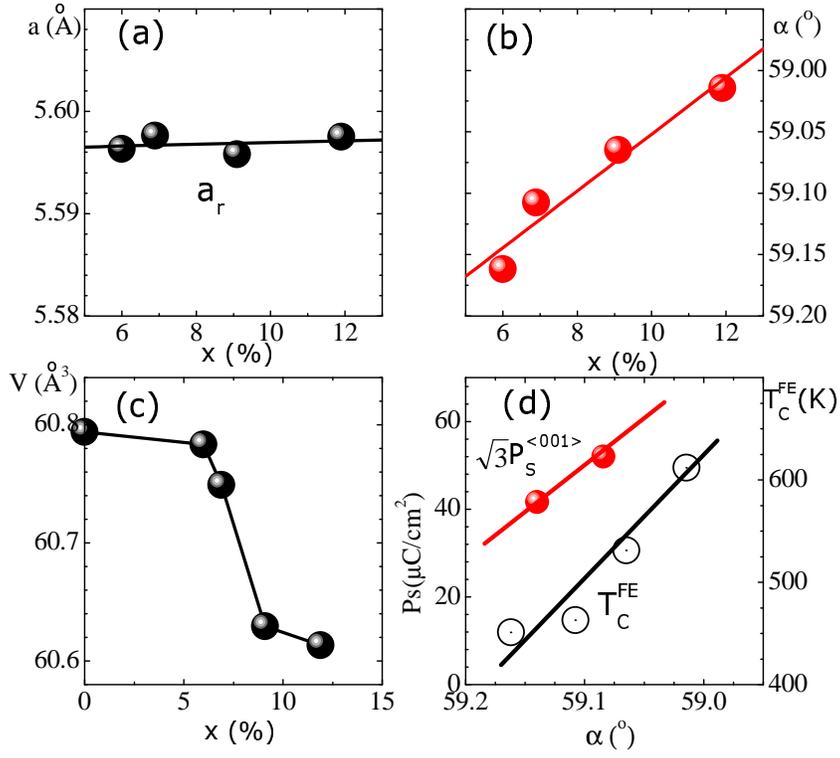}
\caption{\label{fig5} (a) Lattice constants of the rhombohedral cell
for Ag$_{1-x}$Li$_x$NbO$_3$ solid solutions. (b) Change in
rhombohedral angle  $\alpha$. (c) Change in volume of the perovskite
cell. (d) Relationships of ferroelectric phase transition and
spontaneous polarization with $\alpha$. }

\end{figure}

In a recent ${\it ab}$ initio study of AgNbO$_3$, it is shown that
Ag  has a large displacement (0.50 ${\AA}$) than that (0.22 ${\AA}$)
of  Nb  in a tetragonal structure\cite{Grinberg,Grinberg2},
indicating that Ag has a strong ability to shift to an off-centering
position in AgNbO$_3$. It seems that the weak ferroelectricity and
strong antiferroelectricity in AgNbO$_3$ may be reasonably
attributed to Nb and Ag displacements respectively\cite{fu4}. When
Li is introduced into Ag site, the volume of perovskite cell will
decreases with Li-concentration due to the small ionic radius of
Li$^+$ (0.92 ${\AA}$) with respective to Ag$^+$ (1.28 ${\AA}$). Fig.
\ref{fig5}(c) really shows such variation in  the cell volume.
Apparently, suppression of unit cell will reduce the  space
available for atomic displacements in the structure, and hence the
ferroelectricity of the material. However, the polarization
measurements on single crystals and dielectric measurements show
that both spontaneous polarization and ferroelectric phase
transition temperature increase with Li-concentration. Such an
enhancement of ferroelectricity may be accounted for by the very
strong ability of the off-centering of Li in perovskite\cite{Bilc}.
Moving of Li toward an off-centering position then triggers the
large local displacement of Ag along the same direction, leading to
a ferroelectric state with large polarization in the ALN solid
solutions. This inference is also supported by the Li-concentration
dependence of rhombohedral distortion. As shown in Fig. \ref{fig5},
although the lattice constant $a_{\rm r}$ of rhombohedral cell
remains nearly unchanged, however, the rhombohedral angle $\alpha$
decreases monotonically with Li-concentration, indicating that
increasing the Li-concentration enhances the rhombohedral
distortion. As a result, we observed a linear increase in
ferroelectric phase transition temperature and spontaneous
polarization (Fig. \ref{fig5}(d)) with the rhombohedral angle
$\alpha$.

\subsection{Proposed phase diagram}
\begin{figure}
\includegraphics[width=10cm]{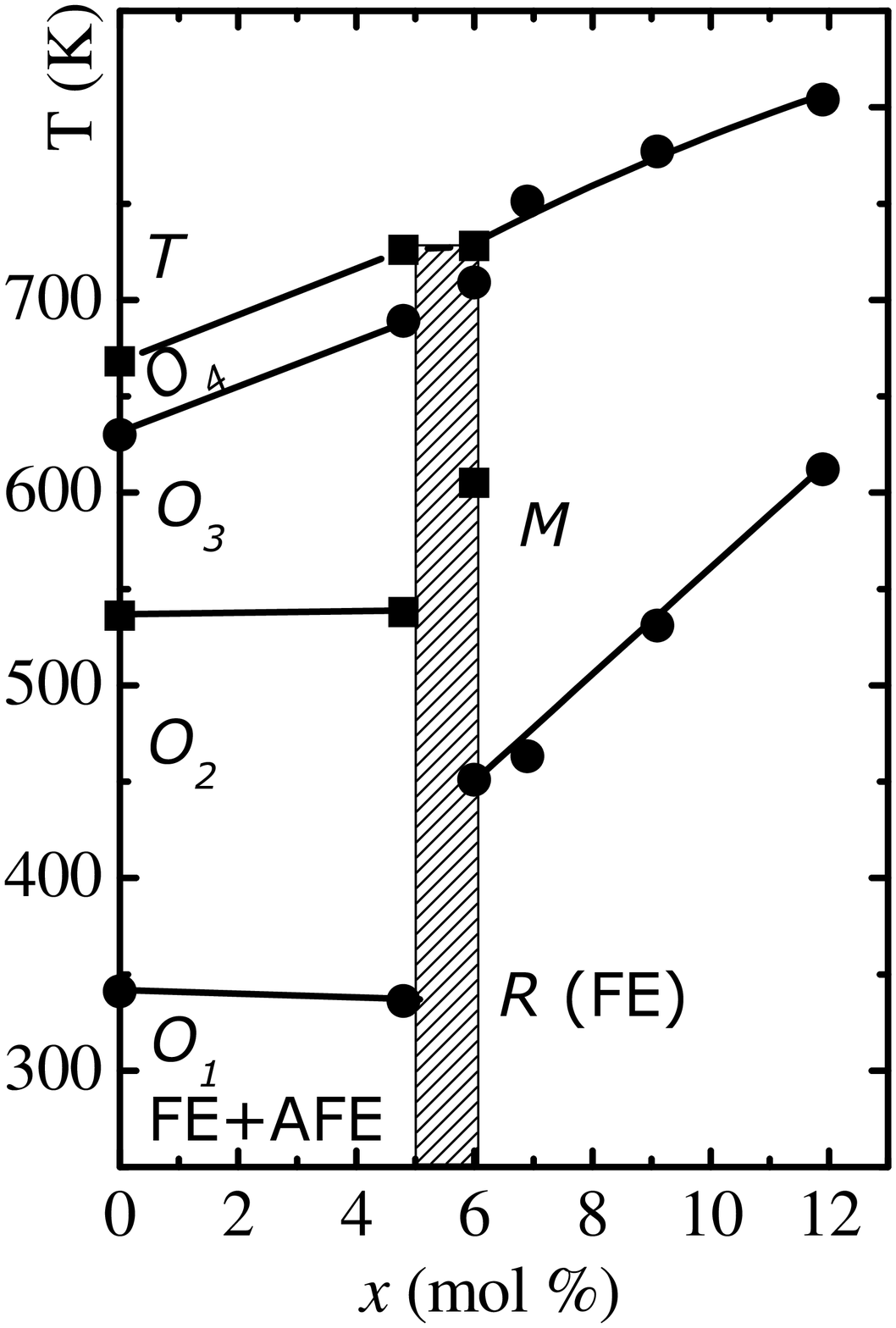}
\caption{\label{fig6} Phase diagram proposed for
Ag$_{1-x}$Li$_x$NbO$_3$ solid solution. Shaded area indicates the
phase boundary. T, O, R, M indicate the tetragonal, orthorhombic,
rhombohedral, and monoclinic structures, respectively.}
\end{figure}

  On the basis of results described above, a phase diagram of ALN
solution is summarized in Fig. \ref{fig6}, where T, O, R and M
represent tetragonal, orthorhombic, rhombohedral, and monoclinic
symmetries, respectively. At room temperature, structure
transformation from O to R phase at $x_{\rm c}$ dramatically changes
the polar nature of ALN. It consists of weak FE and strong AFE in O
phase, but very strong FE in R phase. At higher temperature,
successive phase transitions occur. The exact structures of high
temperature phases remain to be clarified by further investigations
using synchrotron radiation or neutron diffractions. However, the
symmetry may be attributed qualitatively on the basis of the reports
on AgNbO$_3$ \cite{Sciau} and ALN \cite{Nalbandyan,Sakabe}. We
notice that our phase diagram is basically in agreement with those
of Nalbandyan et al. \cite{Nalbandyan} and Sakabe et al.
\cite{Sakabe}, but different from that of Wada et al.\cite{Wada} who
prepared the samples by a process starting from a heavily
non-stoichiometric composition that might cause completely different
phase evolution.

\section{Summary}

In summary, we have demonstrated that strong local polarization of
Ag in AgNbO$_3$ can be effectively induced into a ferroelectric
state by chemically modifying  with  Li substitution, which favors a
rhombohedral distortion. In the rhombohedral phase, the
ferroelectricity is derived from both atomic displacements of A- and
B-site atoms in the perovskite structure. Furthermore, increasing
Li-concentration enhances the rhombohedral distortion, consequently
leading to the linear increase of ferroelectric phase transition and
spontaneous polarization with Li-concentration.

\textbf{Acknowledgments} This work was partly supported by
Collaborative Research Project of Materials and Structures
Laboratory of Tokyo Institute of Technology, and  Grant-in-Aid for
Scientific Research, MEXT, Japan

\clearpage


\clearpage

\begin{table}
\caption{\label{tab1}Structural parameters obtained from Rietveld
refinement for x=0.1. ${\it Note}$:  The isotropic thermal
parameters were kept at a value of 0.5 ${\AA}^2$ in the refinements
because we cannot find significant improvement of the refinement
when changing these parameters.}
  \begin{tabular}{lc}
 \hline
  \hline
  Space group(No.) & R3c(161) \\
  $a/{\AA}$        & 5.52055(9) \\
  $b/{\AA}$        & 5.52055(9) \\
  $c/{\AA}$        & 13.7938(3) \\
  $\alpha$         & 90 \\
  $\beta $         & 90 \\
  $\gamma$          & 120 \\
  volume/${\AA}^3$   & 364.07 \\
  \hline
  \end{tabular}

  \begin{tabular}{ccccc}
   \hline
  Atom &  site  & x & y & z \\
   \hline
  Ag/Li & 6a  & 0 & 0 & 0.2545(8) \\
  Nb    & 6a  & 0 & 0 & 0.0097(8) \\
  O     & 18b & 0.5533 & 1 & 0.2599(9) \\
  \hline
  \hline
 \end{tabular}

 \end{table}


\clearpage

\clearpage

\clearpage

\clearpage

\clearpage

\clearpage


\begin{thebibliography}{30}


\bibitem{Lines}  Lines M E and Glass A M 1977 {\it Principle and Application of Ferroelectrics and Related
materials}, (Oxford: Clarendon Press, )

\bibitem{Jaffe} Jaffe B, Cook  W R Jr and Jaffe H  1971 {\it Piezoelectric ceramics},
(London:Academic)

\bibitem{Scott} Scott J F 2000 {\it Ferroelectric memories}, (Berlin:Springer)

\bibitem{fu0} Fu D, Itoh M, Koshihara S, Kosugi T and Tsuneyuki S 2008 Phys. Rev. Lett. \textbf{100} 227601

\bibitem{Saito}  Saito Y,  Takao H,  Tani T,  Nonoyama T,   Takatori K,  Homma T,   Nagaya T and
Nakamura M 2004 Nature (London) \textbf{432} 84


\bibitem{fu1} Fu D, Itoh M and Koshihara S 2008 Appl.  Phys. Lett. \textbf{93}
012904

\bibitem{fu2} Fu D, Itoh M and  Koshihara S 2010  J. Phys.: Condens. Matter.  \textbf{22} 052204

\bibitem{fu3} Tazaki R, Fu D, Itoh M, Daimon M and Koshihara S 2009 J. Phys.: Condens. Matter \textbf{21} 215903

\bibitem{Bilc} Bilc D I  and Singh D J 2006 Phys. Rev. Lett. \textbf{96} 147602

\bibitem{Grinberg} Grinberg I and Rappe A M 2004 Appl. Phys. Lett. \textbf{85} 1760

\bibitem{Cohen} Cohen R E 1992 Nature (London) \textbf{358} 136

\bibitem{Egami} Egami T,  Dmowski W, Akbas M and  Davies P K 1998 in {\it First-principles Calculations for Ferroelectrics-Fifth Williamsburg
Workshop}, edited by R. Cohen ( NY:American Institute of
Physics),pp. 1-10


\bibitem{Kuroiwa}   Kuroiwa Y et al. 2001 Phys. Rev. Lett. \textbf{87} 217601

\bibitem{Grinberg2}  Grinberg I and  Rappe A M 2003 in {\it Fundamental Physics of Ferroelectrics 2003}, edited by P. K Davies and D. J. Singh ( NY: American Institute of Physics), pp. 130-138

\bibitem{Kato}  Kato H,Kobayashi H and Kudo A 2002 J. Phys. Chem. B \textbf{106}, 12441

\bibitem{Sciau}SciauPh, Kania A, Dkhil B, Suard E and  Ratuszna A 2004 J. Phys.: Condens. Matter \textbf{16} 2795

\bibitem{fu4} Fu D, Endo M, Taniguchi H, Taniyama T and  Itoh M 2007 Appl. Phys. Lett.  \textbf{90} 252907


\bibitem{Francobe}  Francobe M H andLewis G 1958 Acta Crystallogr. \textbf{11} 175

\bibitem{fu5}Fu D, Endo M, Taniguchi H, Taniyama T, Koshihara S and Itoh M 2008 Appl. Phys. Lett.  \textbf{92} 172905

\bibitem{fu6} Fu D, Itoh M and Koshihara S 2009 J. Appl.  Phys. \textbf{106} 104104

\bibitem{Levin} Levin I,  Krayzman V,  Woicik J C,  Karapetrova J,  Proffen T,
Tucker M G and  Reaney I M 2009 Phys. Rev. B \textbf{79} 104113

\bibitem{Pawelczyk}Pawelczyk M 1987 Phase Transitions \textbf{8} 273

\bibitem{Petzelt} Petzelt J,  Kamba S, Buixareras E,  Bovtun V,  Zikmund Z,
Kania A,  Koukal V,  Pokorny J,Polvka J, Pashkov V,  Komandin G and
 Volkov A 1999 Ferroelectrics \textbf{223} 235

\bibitem{Nalbandyan}  Nalbandyan V B, Medviediev B S,  and Beliayev I N 1980 Izv. Akad. Nauk
SSSR, Neorg. Mater. \textbf{16}, 1819 (1980).

\bibitem{Sakabe}  Sakabe Y,  Takeda T,  Ogiso Y and  Wada N 2001 Jpn. J. Appl. Phys.,
Part 1 \textbf{42} 5675

\bibitem{Wada} Wada S,  Saito A,  Hoshina T,  Kakemoto H,  Tsurumi T, Moriyoshi C  and  Kuroiwa Y 2006 Jpn. J. Appl. Phys., Part 1 \textbf{45} 7389

\bibitem{Kania1}  Kania A,  Roleder K and  Lukaszewski M 1984 Ferroelectrics \textbf{52} 265

\bibitem{Kania2} Kania A  and Kwapulinski J 1999 J. Phys.: Condens. Matter \textbf{11} 8933

\bibitem{Shiozaki} {\it Ferroelectrics and Related Substances},Landolt-Bornstein, New Series,
Group III, Vol. 36, Pt. A1, edited by Y. Shiozaki, E. Nakamura, and
T. Mitsui 2001 (Springer, Berlin)

\bibitem{Araujo} C. A. P. de Araujo, J. D. Cuchiaro, L. D. McMillan, M. Scott and J. F. Scott 1995 Nature (London) \textbf{374}, 627


\end{thebibliography}
\end{document}